# Ultrahigh thermal conductivity and strength in direct-gap semiconducting graphene-like BC$_6$N: A first-principles and classical investigation

Bohayra Mortazavi*

*Chair of Computational Science and Simulation Technology, Institute of Photonics, Department of Mathematics and Physics, Leibniz Universität Hannover, Appelstraße 11,30167 Hannover, Germany.*
*Cluster of Excellence PhoenixD (Photonics, Optics, and Engineering–Innovation Across Disciplines), Gottfried Wilhelm Leibniz Universität Hannover, Hannover, Germany*

## Abstract

In recent years, graphene-like boron carbide and carbon nitride nanosheets have attracted remarkable attentions, owing to their semiconducting electronic nature and outstanding mechanical and heat transport properties. Graphene-like BC$_6$N is an experimentally realized layered material and most recently has been the focus of numerous theoretical studies. Interestingly, the most stable form of BC$_6$N monolayer remains unexplored and limited information are known concerning its intrinsic physical properties. Herein, on the basis of density functional theory (DFT) calculations we confirm that the most stable form of BC$_6$N nanosheet shows a rectangular unitcell, in accordance with an overlooked experimental finding. We found that BC$_6$N monolayer is a semiconductor with 1.19 eV HSE06-based direct gap and yields anisotropic and excellent absorption of visible light. First-principles results highlight that BC$_6$N nanosheet exhibits anisotropic and ultrahigh tensile strength and lattice thermal conductivity, outperforming all other fabricated 2D semiconductors. We moreover develop classical molecular dynamic models for the evaluation of heat transport and mechanical properties of BC$_6$N nanomembranes. The presented results in this work not only shed light on the most stable configuration of BC$_6$N nanosheet, but also confirm its outstandingly appealing electronic, optical, heat conduction and mechanical properties, extremely motivating for further theoretical and experimental endeavors.

# 1. Introduction

Graphene [1–3], the most stable form of sp$^2$ carbon atoms in the single-layer form, exhibits extraordinarily high mechanical strength [4], thermal conductivity [5,6] and carrier mobilities [7] and exciting electronic features [8–11]. Graphene since its first successful isolation [1], has kept its position as an ambassador for attracting ongoing and extensive interests toward two-dimensional (2D) materials. As a results of graphene ongoing success, 2D materials has been continuously growing during the last decade. Nature of carbon atoms allows them to form different bonding and evolve to various lattices and yield diverse properties. For example, while graphene is a semimetal with a zero gap, graphdiyne lattices can show metallic, semimetallic or semiconducting characters [12,13] and F-diamane is an insulator [14]. Moreover, although graphene is known to exhibit ultrahigh lattice thermal conductivity, graphdiyne nanosheets show low thermal conductivities [15,16]. Presenting a suitable gap in the electronic structure is a critical necessity for the majority of applications in electronics, optoelectronics and catalysis. The semimetallic nature of pristine graphene thus limits its effectiveness for critical applications in electronics, sensors, energy conversion and optics. Lack of band gap in graphene interestingly stimulated the prediction and synthesis of novel 2D inherent semiconductors, such as transition metal dichalcogenides [17–19], phosphorene [20,21], indium selenide [22] and germanium arsenide [23].

Covalently bonded networks made of carbon and nitrogen atoms, are currently among the most attractive class of 2D materials. Carbon nitride nanosheets are mostly semiconductors and can show highly appealing electronic, optical, mechanical and heat transport properties. The interest toward these lattices originates from outstanding catalytic, electronic and optical [24–28] properties of triazine-based g-C$_3$N$_4$ [29] nanoporous carbon nitrides. In recent years, as a results of extensive experimental endeavors, numerous novel carbon nitride 2D semiconductors have been fabricated successfully, among them we should mention, C$_2$N, so called nitrogenated holey graphene [30], C$_3$N or 2D polyaniline [31], C$_3$N$_{4.8}$, made of a combined triazole/triazine framework [32], all-triazine C$_3$N$_3$ [33], poly(triazine imide) C$_3$N$_4$ [34] and C$_3$N$_5$ made of s-heptazine and azo-linkage [35]. These exciting and continuous experimental successes enhance the interests and application prospect of carbon nitride nanosheets. Nonetheless, among the aforementioned lattices, only the C$_3$N shows a densely packed structure, and the rest include low-density nanoporous lattices. 2D polyaniline C$_3$N is already known to show outstandingly high thermal conductivity [36,37,46,38–45] and mechanical properties [47–51], stemming from its densely packed lattice and strong covalent interactions between carbon and nitrogen atoms. In contrast, the presence of nanoporosity in the atomic structure promotes phonon scattering and can lead to substantial suppression of lattice thermal conductivity, which has been theoretically confirmed [15,52–55]. Interestingly, despite the fact that nanoporous carbon nitride 2D systems may show around three orders of magnitude lower thermal conductivity than graphene [15,52,53], but because of outstanding rigidity of carbon-nitrogen bonds [56–62], they can show lower but comparable tensile strengths to graphene.

Among various carbon nitride 2D structures, recently polyaniline based $C_3N$ has attracted tremendous attention, not only because of its semiconducting nature but also due to its remarkably high thermal transport and mechanical properties. Motivated by the success of $C_3N$ nanosheet, we previously studied the electronic, optical, mechanical and thermal transport properties of graphene-like $BC_6N$ monolayers with different atomic lattices [63]. Interestingly, we found that $BC_6N$ monolayers not only show direct gap semiconducting nature, but also considerably higher lattice thermal conductivity than $C_3N$ counterpart. These highly attractive findings, stimulated numerous theoretical studies on the exploration of electronic properties and effectiveness of $BC_6N$ nanosheets for various applications [64–70]. In our original theoretical investigation [63], we found that among the considered hexagonal lattices, the one that includes boron-nitrogen bonds is energetically most stable. Nevertheless, it appears that $BC_6N$ nanosheet can take a rectangular unitcell, which has been vastly overlooked in previous studies. Worthy to note that $BC_6N$ layered materials have been already fabricated more than a decade ago [71,72] and such that motivating theoretical findings can stimulate further experimental achievements. In this work our objective is to examine the energetic and dynamical stability and electronic properties of different $BC_6N$ lattices on the basis of first-principles density functional theory calculations. We next elaborately explore the optical, mechanical and heat transport properties of the most stable lattice. On the basis of insights provided by first-principles results, we also develop classical molecular dynamic models for the evaluation of heat transport and mechanical properties of $BC_6N$ nanomembranes.

2. Computational methods

Density functional theory (DFT) calculations were performed with the generalized gradient approximation (GGA) and Perdew–Burke–Ernzerhof (PBE) [73] using the *Vienna Ab-initio*

*Simulation Package* (VASP) [74,75]. Projector augmented wave method was used to treat the electron-ion interactions [76,77] with a plane wave cutoff energy of 600 eV and convergence criteria of $10^{-5}$ eV for the energy self-consistent loop. For the geometry optimizations, atoms and lattices are relaxed using conjugate gradient algorithm till the Hellman-Feynman forces drop to lower than 0.001 eV/Å [78]. Periodic boundary conditions were applied in all directions with a 20 Å vacuum layer to avoid image-image interactions along the monolayers' thickness. The first Brillouin zone (BZ) was sampled with 23×23×1 a Monkhorst-Pack [79] k-point grid. Mechanical properties are examined by conducting uniaxial tensile loading simulations. Since PBE method underestimates the conduction band maximum positions, we utilized HSE06 hybrid functional [80] to more accurately evaluate the electronic and optical properties. Light absorption is reported on the basis of frequency-dependent dielectric matrix, constructed over HSE06 results [81]. The mechanical and lattice thermal conductivity are reported by assuming a thickness of 3.35 Å for all considered monolayers.

In order to acquire $2^{nd}$ order harmonic force constants, density functional perturbation theory (DFPT) calculations were carried out using the VASP for 4×4×1 and 8×2×1 supercells for the hexagonal and rectangular unitcells, respectively. Phonon dispersion relations and harmonic force constants are obtained using the PHONOPY code [82]. Moment tensor potentials (MTPs)[83] are trained to evaluate the phononic properties [84] using the MLIP package [85]. Ab-initio molecular dynamics (AIMD) simulations are conducted with a time step of 1 fs over 5×2×1 supercell for the rectangular $BC_6N$ [84]. The training data for the development of MTPs are prepared by conducting three separate AIMD simulations at 50, 700 and 1200 K for 1000 time steps. From every separate calculation, 200 trajectories were subsamples and used for the training. Anharmonic $3^{rd}$ order interatomic force constants for the rectangular $BC_6N$ are obtained for 8×2×1 supercells using the trained MTPs by considering the interactions with eleventh nearest neighbors. Full iterative solution of the Boltzmann transport equation (BTE) with consideration of isotope scattering (naturally occurring samples) is carried out to estimate the phononic thermal conductivity using the ShengBTE [86] package with $2^{nd}$ and $3^{rd}$ order force constants as inputs, as explained in our earlier study [87].

Classical molecular dynamics simulations are conducted using the LAMMPS [88] package. For the evaluation of mechanical response and thermal conductivity, we used time increments of 0.25 and 1 fs, respectively. Before applying the loading conditions, all structures were equilibrated using the Nosé-Hoover barostat and thermostat method (NPT). In accordance with first-principles modelling, mechanical responses are evaluated by conducting the uniaxial tensile simulations. In order to minimize the effects of the loading strain rate, strain is applied with a fixed step of 0.005 and after straining the structures were relaxed to reach negligible stress along the perpendicular direction of loading using the NPT ensemble for 50 ps. We performed non-equilibrium molecular dynamics (NEMD) simulations to examine the thermal conductivity. After equilibration, atoms at the two ends were fixed and system was divided into 22 slabs, and a temperature difference of 20 K was applied using the NVT method between two boundary slabs, while no thermostat was set for the remaining 20 slab. The lattice thermal conductivity was predicted on the basis of 1D Fourier's law, by calculating the applied heat flux by the NVT and established averaged temperature gradient along the samples' length.

3. Results and discussions

We first study the structural and stability of considered graphene-like monolayers. Fig. 1a displays the crystal structure of experimentally realized $BC_3$ and $C_3N$ sheets and $BC_6N$ 2D crystals with four different atomic configurations. All geometry optimized structures are included in the supplementary information document. The lattice constant of $BC_6N$ monolayers with hexagonal unitcells are found to be very close and around 4.98 Å, which is in between of those for $C_3N$ (4.86 Å) and $BC_3$ (5.17 Å) monolayers. For every structure, we also include the corresponding energy per atom value on the basis of PBE functional. Results shown in Fig. 1 clearly reveal that the rectangular lattice of $BC_6N$ exhibits the minimum energy and is thus the most stable structure. In fact, this finding is in an excellent agreement with an overlooked experimental observation [72]. Interestingly, the $BC_6N$ lattice with rectangular unitcell exhibits lower energy than $BC_3$ and $C_3N$ monolayers. The higher stability of $BC_6N$ lattice with rectangular unitcell compared to $BC_3$ and $C_3N$ monolayers can be explained in terms of different amount of π-conjugation in these three monolayers. As shown in Fig. 1, in both $BC_3$ and $C_3N$ monolayers, isolated carbon hexagons are surrounded either by boron or nitrogen atoms, respectively. In an earlier work on the nature of chemical bonds in $BC_3$ monolayer, Papove and Boldyrev [89] showed that carbon hexagons exhibit benzene-like π-bonding while other hexagons made of carbon and boron are anti-aromatic. In another work, Tan *et al.* [90] found that in $C_3N$ monolayer also π-electrons are mostly localized on carbon hexagons with a minor extension over the nitrogen atoms. The situation in $BC_6N$, however, is different than $BC_3$ and $C_3N$ monolayers. The crystal structure of $BC_6N$ consists of two rings wide zigzag-edge graphene nanoribbons connected by polyacetylene-like BN chains. From chemistry, we know that π-conjugation in a system lowers its energy and increases the stability. The amount of conjugation in BC$_6$N is clearly much larger than that in BC$_3$ and C$_3$N monolayers, resulting in its higher stability. This also explains the higher stability of BC$_6$N with rectangular unitcell in comparison with those with hexagonal unitcells. On the other side, we should also note that the energy per atom of graphene and hexagonal boron-nitride monolayers with PBE functional are around -9.225 and -8.799 eV, respectively. It is clear that since BC$_6$N with rectangular unitcell intuitionally resembles more to graphene, it is energetically more stable than other counterparts in which boron and nitrogen atoms are periodically distributed in their lattices. In Fig. 1, we also plot electron localization function (ELF) [91] to investigate the nature of chemical bonds in these monolayers. ELF is a topological function and takes a value between 0 to 1, which quantitatively differentiate between different types of chemical bonds. ELF values close to 1 indicate strong covalent interaction or lone pair electrons, while lower values correspond to metallic or ionic bonds, or weak Van der Waals interactions. It is noticeable that for all considered 6 monolayers, around the center of bonds the ELF values are more than 0.8, which as expected confirm the covalent bonding in these systems. The structural and energetic parameters of studied monolayers are summarized in Table 1.

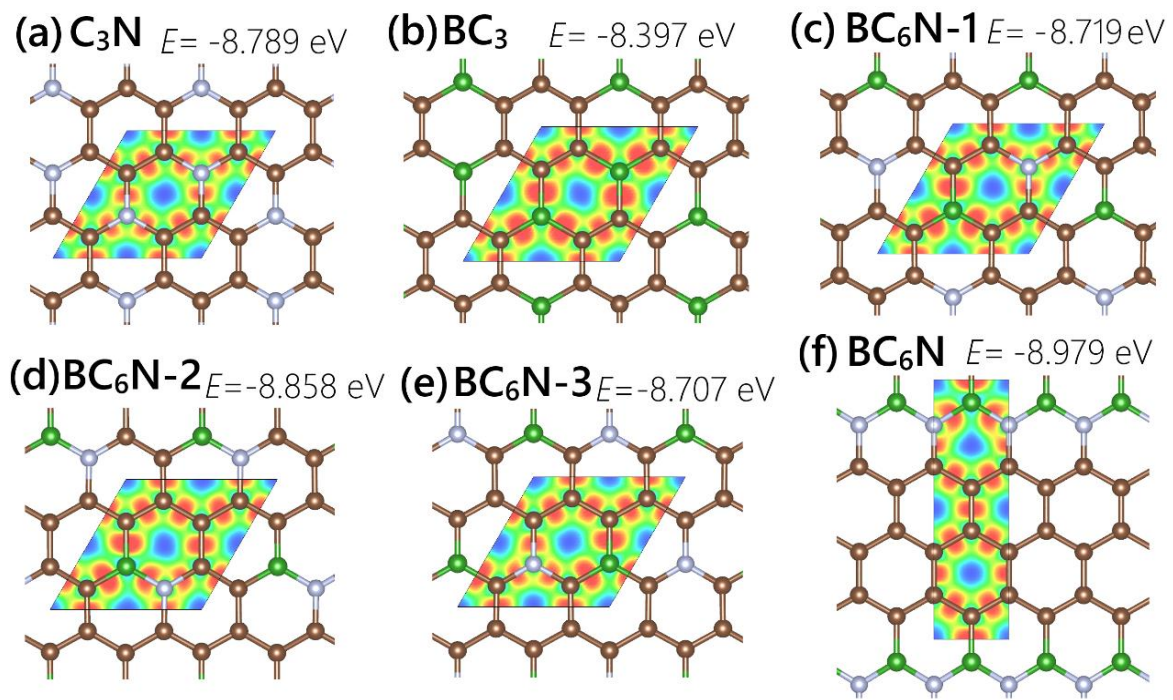

**Fig. 1**, Top views of geometry optimized monolayers along with the corresponding energy per atom on the basis of PBE/GGA functional. Contours illustrate electron localization function, which ranges from 0 (blue) to 0.9 (red). VESTA package was used to prepare this illustration [92].

Table 1, Calculated structural, energetic and electronic properties of considered monolayers. For the band gap the superscripts, D and ID, respectively stand for direct and indirect.

|  | $BC_6N$ | $C_3N$ | $BC_3$ | $BC_6N$-1 | $BC_6N$-2 | $BC_6N$-3 |
|---|---|---|---|---|---|---|
| Energy (eV/atom) | -8.979 | -8.789 | -8.397 | -8.719 | -8.858 | -8.707 |
| Lattice constant (Å) | $2.474^{zig}$, $8.644^{arm}$ | 4.861 | 5.173 | 4.978 | 4.972 | 4.991 |
| PBE gap (eV) | $0.71^D$ | $0.39^{ID}$ | $0.66^{ID}$ | $1.32^D$ | $1.05^D$ | $0.21^D$ |
| HSE06 gap (eV) | $1.19^D$ | $1.06^{ID}$ | $1.85^{ID}$ | $2.07^D$ | $2.72^D$ | $0.77^D$ |

After systematically finding the most stable $BC_6N$ lattice in the 2D form, we next examine the dynamical stability by calculating the phonon dispersion relations. The acquired phonon dispersion relations by the DFPT method for the considered monolayers along BZ are illustrated in Fig. 2. Considered monolayers show three acoustic modes starting from the Γ point, two with linear dispersions and the other one with quadratic relation (ZA mode). Moreover, optical modes for all lattices extend to frequencies around the 45 THz. Calculated phonon dispersions confirm the dynamical stability of considered monolayers, because they do not exhibit imaginary frequencies. As an interesting preliminary finding, it is clear that acoustic modes in the most stable $BC_6N$ lattice, particularly the ZA mode, show considerably wider dispersions than corresponding modes in other considered lattices. This implies higher phonon's group velocities of these modes, which can enhance the thermal transport.

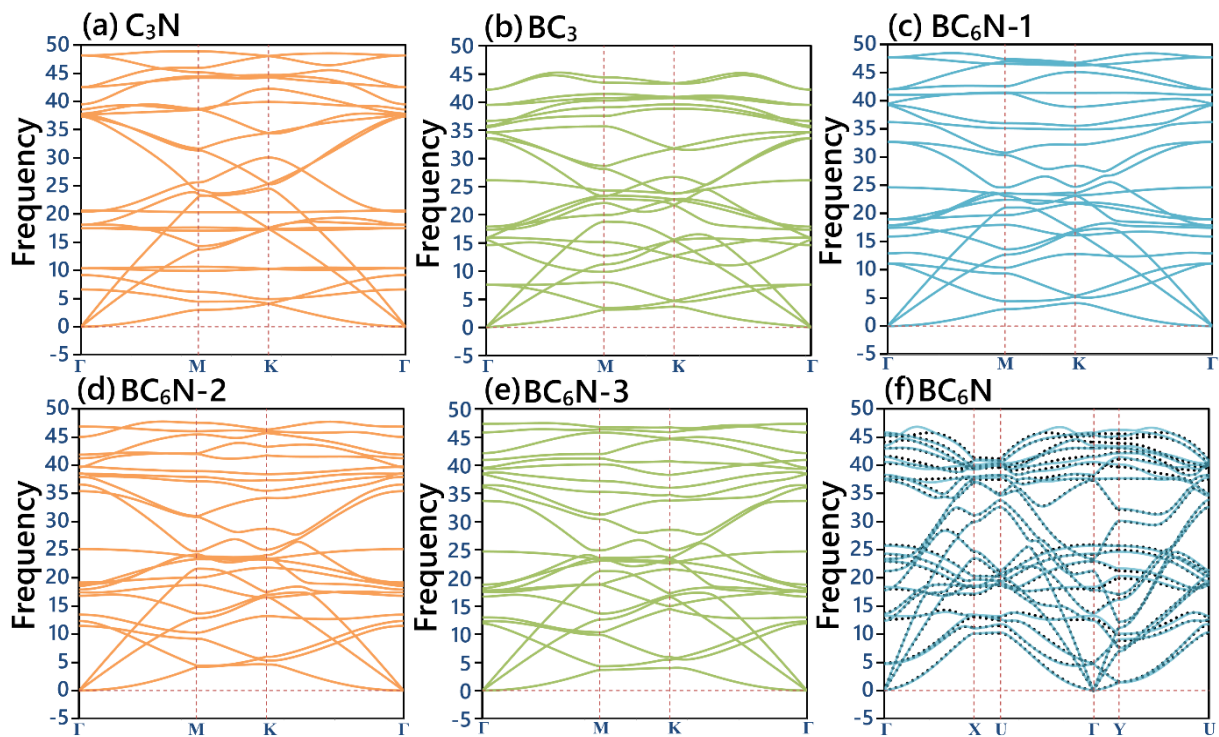

Fig. 2, Phonon dispersion relations predicted by the DFPT method. For the rectangular $BC_6N$ monolayer (panel f) the results with fitted MTP are also plotted with dotted lines.

We next shift our attention to explore electronic properties of considered BC$_6$N monolayers and compare the findings with those of BC$_3$ and C$_3$N counterparts. Fig. 3 depicts the electronic band structures acquired using the PBE and HSE06 methods. From the presented results, it is apparent that the electronic structure varies considerably with the atomic configurations. As expected, the general features of PBE and HSE06 band structures are consistent. As an interesting finding, while BC$_3$ and C$_3$N monolayers are indirect gap semiconductors, considered four BC$_6$N monolayers are direct gap semiconductors. It is found that most stable BC$_6$N monolayer exhibits a direct gap of 0.71 and 1.19 eV according to the PBE and HSE06 methods, respectively. The electronic band gap of BC$_6$N monolayer with rectangular unitcell predicted by the more accurate HSE06 functional is closer to that of the C$_3$N monolayer, 1.06 eV and distinctly lower than that of the BC$_3$ counterpart, 1.85 eV. For the most stable BC$_6$N monolayer, it is conspicuous that the valance band maximum (VBM) and conductance band maximum (CBM) predicted by both PBE and HSE06 band structures occur in between the Γ-X path. For C$_3$N monolayer the VBM and CBM occur at M and Γ points of BZ, respectively, reversed as those of BC$_3$ lattice, in which VBM and CBM happen at Γ and M points of BZ, respectively. Interestingly, despite of very different atomic configurations and electronic band gaps, for the BC$_6$N monolayers with hexagonal unitcells, both VBM and CBM occur in the vicinity of K point in BZ. The predicted electronic band gaps by the PBE and HSE06 methods for the considered monolayers are summarized in Table 1. We particularly investigated the effects of uniaxial tensile straining along the armchair and zigzag directions on the evolution of electronic band gap of BC$_6$N monolayer and the results are shown in Fig. S1. We found that for the uniaxial straining along the zigzag direction, the band gap first slightly increases but at higher strain levels decreases continuously. For the uniaxial straining along the armchair direction the electronic gap decreases almost linearly with strain. These preliminary results highlight that BC$_6$N monolayer shows strain-tunable electronic gap, which can be appealing for further explorations.

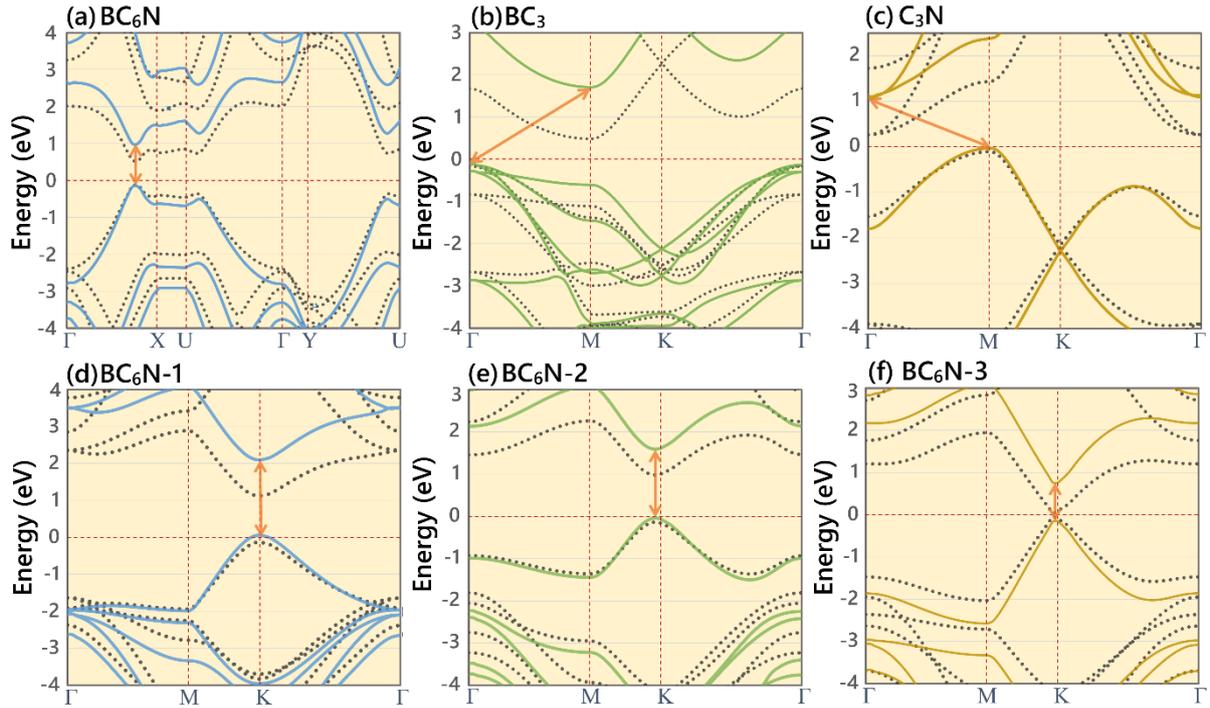

Fig. 3, Electronic band structures predicted by PBE (dotted lines) and HSE06 (continues lines). The arrow indicates the position of band gap.

Direct and narrow band gap of the most stable $BC_6N$ monolayer are promising indications for its performance in optoelectronic applications. In order to investigate this prospect we used HSE06 functional to calculate light absorption coefficients along the armchair and zigzag directions [93–96] and the obtained results are shown in Fig.4. In this case, we also compare the results with those calculated for $BC_3$ and $C_3N$ monolayers. While $BC_3$ and $C_3N$ monolayers exhibit isotropic absorption coefficients in response to the light polarized along zigzag and armchair directions, $BC_6N$ monolayer in contrast yields strong anisotropic absorption coefficients. The first absorption peak of $BC_6N$, $BC_3$ and $C_3N$ monolayers appear in the visible range, but stemmed from the narrower and direct electronic gap, the first peak for $BC_6N$ lattice occurs at distinctly lower energies. It is noticeable that all three first peaks for $BC_6N$ monolayer along the armchair and zigzag occur in the visible range of light. Along the zigzag direction, the light absorption is higher than armchair direction, which pronounces substantially for the second and third peaks. It can be seen that $BC_6N$, $BC_3$ and $C_3N$ monolayers show remarkably high absorption coefficients ($10^5$ cm$^{-1}$) in a broad range of visible light frequencies, which are comparable to those of halide perovskites [97]. With a direct band gap and large absorption coefficients in visible range, $BC_6N$ monolayer can be a promising candidate for nanoelectronics and optoelectronic applications. Particularly, highly anisotropic optical absorption of $BC_6N$ nanosheet can be extremely appealing for the employment in angle-dependent devices, such as polarized lasers and sensors, photodetectors and digital inverters. Presented results show highly attractive electronic and optical characteristics of BC$_6$N nanosheet, which highlight the need for more elaborated and in-depth investigations.

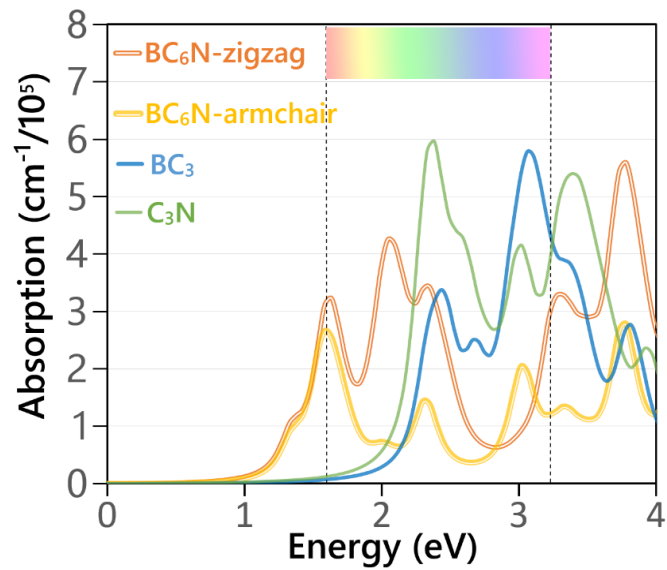

**Fig. 4**, Light absorption spectra of most stable BC$_6$N, BC$_3$ and C$_3$N monolayers calculated using HSE06 functional. The visible-light energy range is also shown by vertical lines.

We next elaborately examine the phononic thermal transport in BC$_6$N monolayer. In order to provide a more useful vision, we also compare the results with that predicted for the pristine graphene using the similar computational procedure. In our calculations for BC$_6$N we considered 8×2×1 supercells with 128 atoms in evaluating the 2$^{nd}$ and 3$^{rd}$ order interatomic force constants. For the sake of consistency, for the graphene monolayer we also selected 8×8×1 supercells with 128 atoms in our calculations. In Fig. 2f for the BC$_6$N monolayer, the MTP-based phonon dispersion (shown with dotted line) is compared with that acquired using the DFPT method, which reveals excellent agreements, particularly for acoustic modes. The temperature dependent lattice thermal conductivity of graphene and BC$_6$N monolayers are plotted in Fig. 5, assuming a thickness of 3.35 Å. For graphene the lattice thermal conductivity is found to be isotropic, whereas BC$_6$N monolayer clearly yields an anisotropic lattice thermal conductivity. By taking into account the isotope scattering, the room temperature lattice thermal conductivity of naturally occurring graphene and BC$_6$N monolayers along the zigzag and armchair directions are predicted to be 3640, 1890 and 1130 W/m.K, respectively. The predicted thermal conductivity of graphene monolayer at 300 K is in an excellent agreement with previous full-DFT BTE reports of 3550 [41], 3845 [98], 3720 [99], 3590 [100], and 3288 W/m.K [101]. Interestingly, due to the presence of boron-nitrogen chains, the thermal conductivity along the zigzag and armchair direction reduces by around 49 and 69 %, respectively, as compared with that of the graphene. Along the armchair direction the boron-nitrogen chains are directly perpendicular to the heat transfer and yield maximal phonon scattering, resulting in a lower thermal conductivity than the zigzag direction, in which the aforementioned chains do not directly face the heat transport direction. As a general rule, phononic thermal conductivity decreases by temperature following a $\sim T^{-\alpha}$ trend, where α is the temperature power factor. The temperature power factors for the thermal conductivity of graphene and $BC_6N$ along the zigzag and armchair directions are predicted to be 1.36, 1.18 and 1.1, respectively. [102]. The smaller α value for $BC_6N$ nanosheet reveals lower Umklapp scattering in this lattice than graphene. Worthy to mention that the lattice thermal conductivity of $C_3N$ and $BC_3$ monolayers with full-DFT BTE solution were predicted to be in the ranges of 80-482 [103–106] and 278-410 W/m.K [63,103], respectively. As it is clear, $BC_6N$ monolayer shows severalfold higher lattice thermal conductivity than $C_3N$ and $BC_3$ counterparts. The outstandingly high thermal conductivity of $BC_6N$ monolayer can be intuitively attributed to the existence of graphene nanoribbons, which facilitate the heat transport, particularly along the zigzag direction. Our results highlight the outstandingly high thermal conductivity of $BC_6N$ nanosheet, which remarkably records the highest value among all produced semiconducting 2D materials.

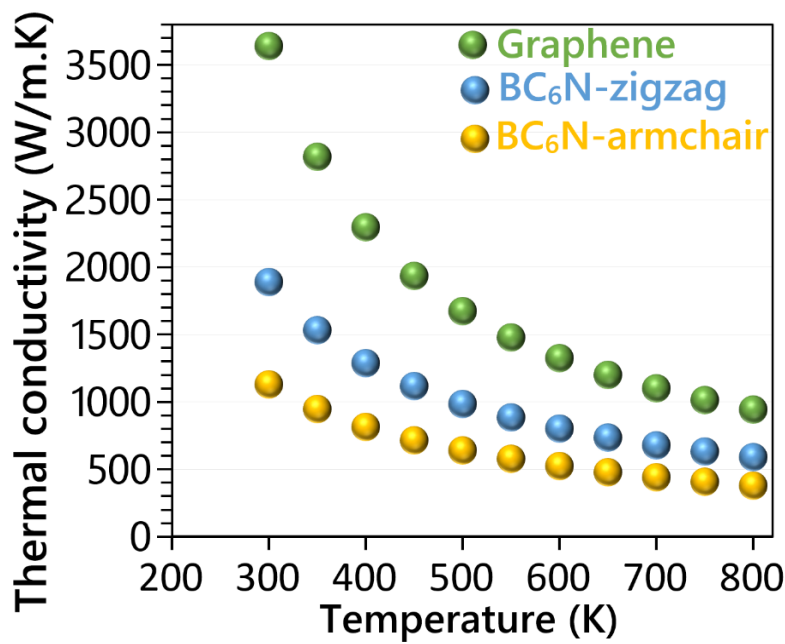

**Fig. 5**, Temperature dependent lattice thermal conductivity of $BC_6N$ monolayer along the armchair and zigzag directions predicted by the BTE solution of lattice thermal conductivity. For the comparison, the corresponding results for graphene are also included.

We next discuss the modeling of thermal transport along the and BC$_6$N nanosheets using the classical molecular dynamics simulations. Optimized Tersoff potential proposed by Lindsay and Broido [107] is currently the most accurate interatomic potential to simulate thermal transport in 2D carbon structures. Since BC$_6$N lattice also includes boron and nitrogen atoms, in order to simulate this system, as a common approach two different Tersoff potential parameter sets can be combined using the mixing rule proposed in the original work by Tersoff [108]. Because of the identical atomic structure, bonding mechanism and similar phonon dispersion relations of h-BN and graphene, we employ two different Tersoff potentials developed for h-BN nanosheet, proposed by Lindsay and Broido [109] and Kinarci *et al.* [110] and accordingly set the Tersoff interatomic potentials for BC$_6$N lattice. As discussed earlier, in the NEMD method atoms at two boundaries are fixed, which certainly limits the wave length of contributing phonon modes. In order to evaluate the single-layer BC$_6$N diffusive lattice thermal conductivity, we carried out NEMD simulations for samples with different lengths to assess the length effect. In Fig. 6, NEMD results for the length effect on the thermal conductivity of BC$_6$N monolayer using the two different interatomic potentials at room temperature are plotted. In this case, we include results along the armchair and zigzag directions and also for the comparison the corresponding results for graphene are also included (find Fig.1a inset). As expected for samples with low lengths, the thermal conductivity increases sharply, presenting ballistic thermal transport. By further increase of the length the enhancement of lattice thermal conductivity decrease and the heat transfer approaches the diffusive regime. As a common approach by extrapolation of the NEMD results for the samples with various lengths, $k_L$, the diffusive phononic thermal conductivity of BC$_6$N monolayer, $k_\infty$, can be calculated, using the following relation [111,112]:

$$\frac{1}{k_L} = \frac{1}{k_\infty}\left(1 + \frac{\Lambda}{L}\right) \qquad (1)$$

where, $\Lambda$ is the effective phonon mean free path. By employing the aforementioned fitting approach, diffusive lattice thermal conductivities were predicted (the fitted curves are plotted in Fig.6). For the single-layer graphene, we predicted the diffusive thermal conductivity of 2690 ±80 W/m.K, in an excellent agreement with previous molecular dynamics results, 2700±80 W/m.K [113,114]. Based on the Lindsay and Broido [109] interatomic potential, the diffusive lattice thermal conductivity of BC$_6$N monolayer along the zigzag and armchair directions at room temperature are predicted to be to be 1150±30 and 660±20 W/m.K, respectively. Using the interatomic potential proposed by Kinarci *et al.* [110], corresponding values of 1080±30 and 470±20 W/m.K, respectively, were predicted. Along the armchair direction, the thermal conductivity of BC$_6$N monolayer is around 60, 57 and 44% of that along the zigzag direction, according to the first-principles results, and classical models by Lindsay and Broido [109] and Kinarci *et al.* [110], respectively. It is thus conspicuous that the Tersoff-based classical model by Lindsay and Broido [109] can more accurately reproduce the lattice thermal conductivity of BC$_6$N monolayer. The employed Tersoff potential for the modeling of thermal transport in graphene or BC$_6$N nanosheet is included in the supplementary information document.

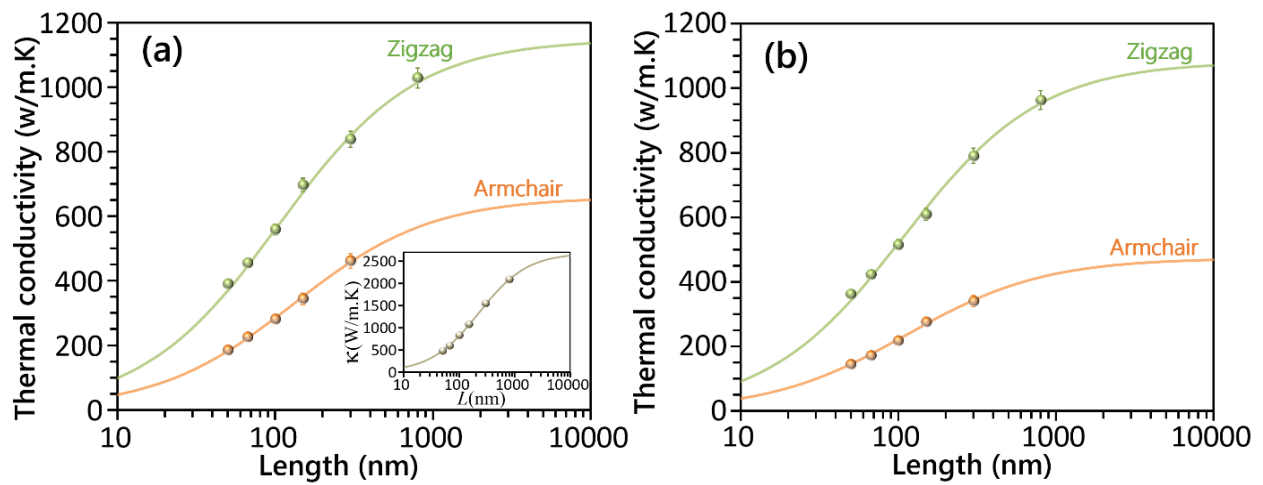

Fig. 6, Classical molecular dynamics results for the length effect on the thermal conductivity of single-layer BC$_6$N at 300 K along the armchair and zigzag directions using the Tersoff potentials by (a) Lindsay and Broido [109] and (b) Kinarci *et al.* [110] . The corresponding results for graphene are shown in inset for panel a. The solid lines illustrate the fitted curves to the NEMD data points.

To better understand the underlying mechanism resulting in reducing the thermal conductivity of BC$_6$N monolayer in comparison with graphene, the phonon's life time and group velocity of these two lattices are compared in Fig. 7. As it can be seen from the phonon dispersion of BC$_6$N monolayer, because of lower symmetry more phonon bands appear and they also cross each other frequently, which surge the scattering and accordingly reduce the phononic thermal conductivity. This is also conspicuous from phonon's life time, shown in Fig. 7a, which reveals distinctly suppressed life time for phonon modes in BC$_6$N monolayer than graphene. From the basics, wider dispersion for a phonon mode reveals its faster group velocity, which can result in a higher thermal conductivity. From the phonon dispersion results shown in Fig. 2 it is clear that BC$_6$N monolayer with rectangular lattice exhibits wider acoustic modes than BC$_3$ and C$_3$N and hexagonal BC$_6$N monolayers, which is illustrative of its higher phonon's group velocities. It is also noticeable that along the Γ-X and Γ-U paths of BZ, the ZA mode show considerably wider frequency than that along the Γ-Y path of BZ, which can be attributed due to the direct interference with boron-nitride chains along the armchair (Γ-Y) direction. In comparison with graphene, BC$_6$N monolayer nonetheless show lower phonon's group velocities as shown in Fig. 7b. Analysis of each phonon mode contribution reveal that ZA acoustic mode is the dominant heat carrier in BC$_6$N monolayer and yields around 67% of total lattice thermal conductivity. Along the armchair and zigzag directions, we found that optical modes contribute marginally to total lattice thermal conductivity, around 17 and 11%, respectively.

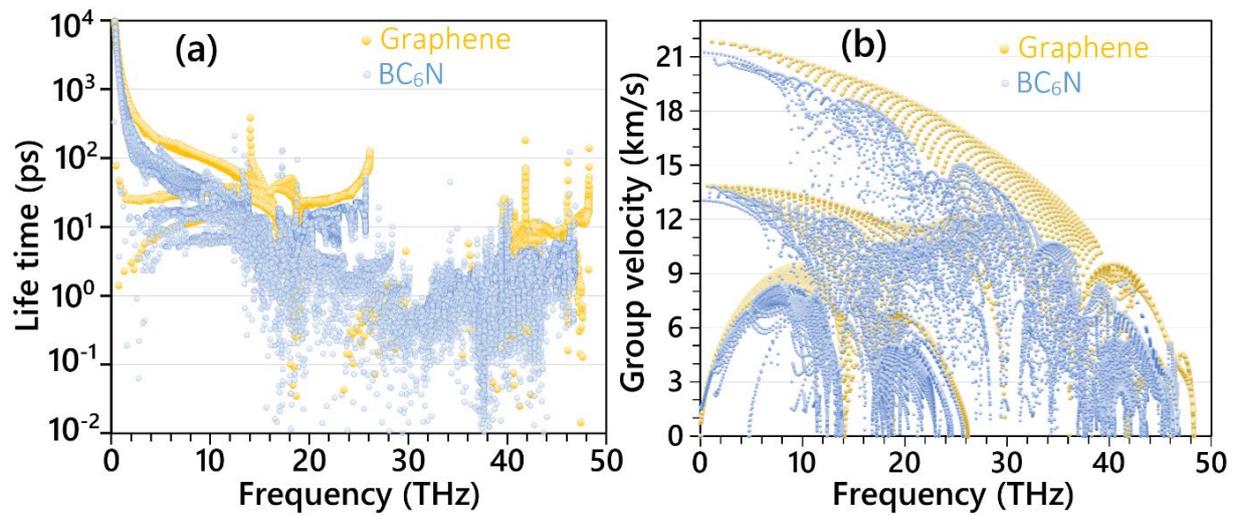

Fig. 7, Phonon's (a) life time and (b) group velocity in graphene and BC$_6$N monolayers predicted by the BTE solution of lattice thermal conductivity.

We finally investigate the mechanical properties of BC$_6$N monolayer by evaluating the uniaxial stress-strain relations and compare the acquired results with graphene. In these calculations, the stresses along the two perpendicular directions of the loading are ensured to stay negligible during various stages of the loading. Due to the contact with vacuum along the normal direction of the monolayers, the stress along this direction automatically reaches to a negligible value upon the geometry optimization. For the other planar direction, the periodic box size is altered to ensure the negligibility of stress. For BC$_6$N and graphene monolayers the mechanical responses are evaluated along the armchair and zigzag directions to examine the anisotropicity. The predicted uniaxial stress-strain relations by the DFT method are illustrated in Fig. 8. The uniaxial stress-strain curves start with a linear relation associated with the linear elasticity, followed by a nonlinear trend up to the maximum tensile strength point. Graphene is found to show an isotropic elastic modulus of 1000 GPa, independent of the loading direction. In accordance with our earlier observations for optical absorption and lattice thermal conductivity, BC$_6$N monolayer exhibits anisotropic mechanical properties, with elastic modulus of 945 and 920 GPa for the loading along the zigzag and armchair directions, respectively. It is conspicuous that BC$_6$N monolayer shows very close tensile strength to graphene for the loading along the zigzag direction, with only around 2 GPa difference. The tensile strength of BC$_6$N (graphene) along the zigzag and armchair are found to be 111.1(113.3) and 87.5(102.7) GPa, respectively. Similar to our results for thermal transport, it is clear that for the loading along the armchair direction the boron-nitrogen chains yield considerable weakening effect on the tensile strength, as they are directly perpendicular to load transfer. In Fig. 8 insets we also compare the structures at the maximum tensile strength point along with their corresponding ELF contours. It is observable that for the loading along the zigzag direction, boron-nitrogen bonds show identical behavior to their parallel carbon-carbon bonds. Nevertheless, in comparison with carbon-carbon bonds, for the boron-nitrogen counterparts the ELF values are slightly lower, which explain the marginally lower tensile strength of BC$_6$N than graphene for the loading along the zigzag direction. In contrast for the loading along the armchair direction, it is conspicuous that boron-carbon bonds extend considerably larger than other bonds in this system, resulting in a significant decline of the tensile strength. In other words, it can be concluded that boron-carbon bonds are the main weakening interactions in the BC$_6$N nanosheet and the failure is expected to occur along these bonds when loaded along the armchair direction.

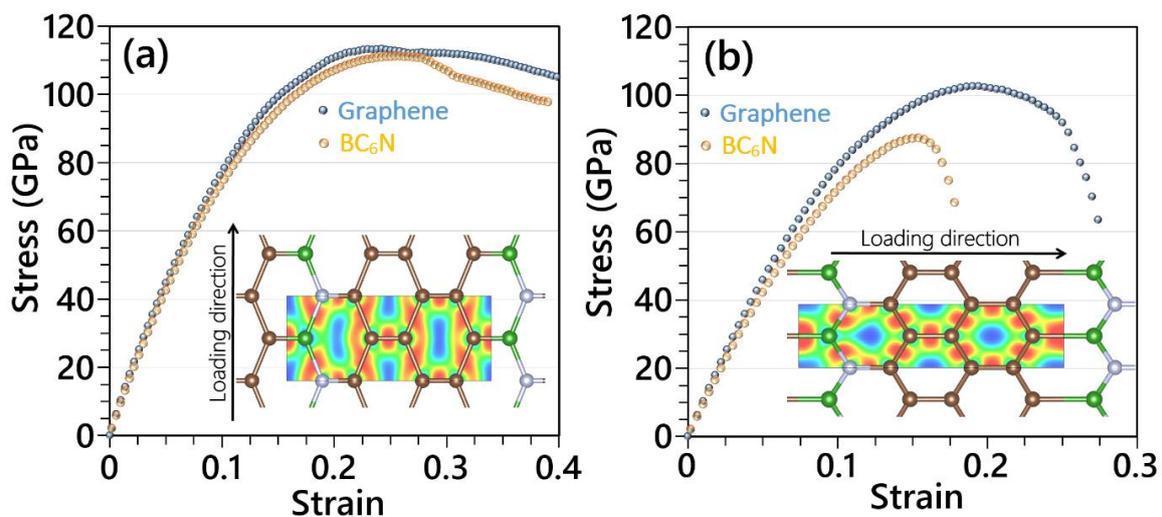

**Fig. 8**, Uniaxial stress-strain curves of graphene and BC$_6$N monolayers along the (a) zigzag and (b) armchair directions predicted by DFT method. Insets illustrate the structures at the maximum tensile strength point along with ELF contours, ranging from 0 (blue) to 0.9 (red).

We next investigate the modeling of mechanical properties of BC$_6$N monolayer using the classical molecular dynamics simulations. As discussed in our earlier study [114], using the original parameter set proposed by Lindsay and Broido [31], the calculated stress–strain curve of pristine graphene at room temperature shows an unphysical strain hardening at high strain levels. Similar artifact has been also reported for the modeling of mechanical properties of graphene by AIREBO and REBO potentials [115]. To resolve this issue, the most common and widely used recipe is to modify the cutoff value of the AIREBO potential [116]. In a similar procedure, in the Tersoff potential for graphene [31], a cutoff function is defined for interactions between 0.18 and 0.21 nm. In accordance with AIREBO potential, we found that by changing the aforementioned cutoff distance to 0.20 and 0.21 nm the unphysical strain hardening in the stress-strain relation can be removed and the modified potential can reproduce the tensile strength of around 130 GPa for pristine graphene [114]. Nevertheless, in order to reproduce the tensile strength of around 110 GPa for graphene predicted by the DFT, we found that the cutoff distance should be further altered to 0.205 and 0.21 nm. In Fig. 9 the uniaxial stress-strain relation of graphene and BC$_6$N monolayers at the room temperature predicted by the classical molecular dynamics simulations are plotted, which shows closer agreements with first-principles results reported earlier (see Fig. 8).

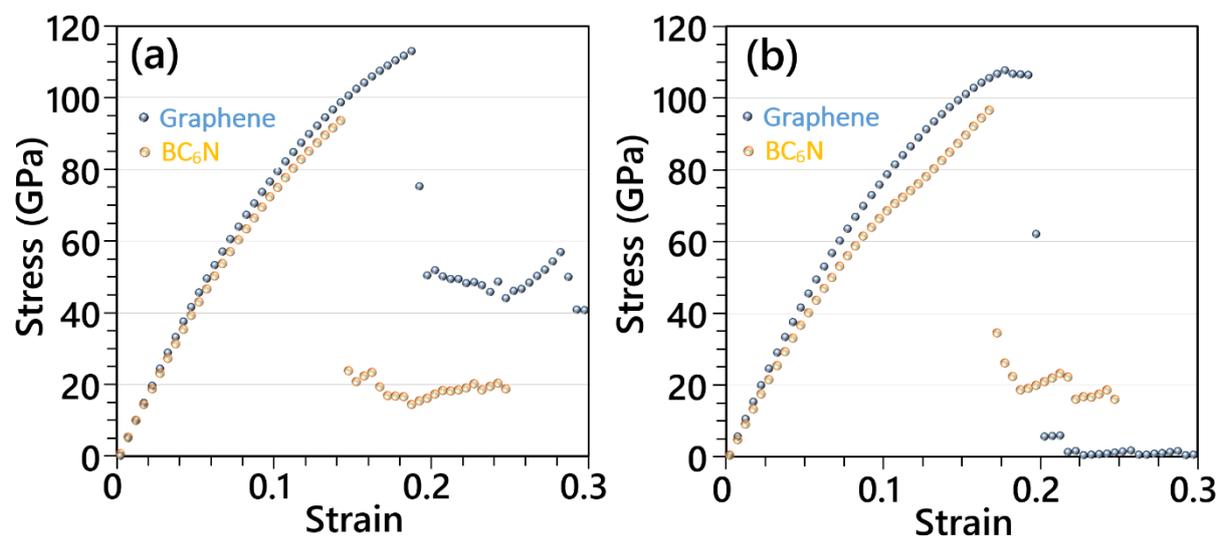

**Fig. 9**, Uniaxial stress-strain curves of graphene and BC$_6$N monolayers along the (a) zigzag and (b) armchair directions at the room temperature predicted by classical molecular dynamics simulations.

For the modeling of mechanical properties in BC$_6$N monolayer using the classical molecular dynamics simulations, employment of Lindsay and Broido [109] potential shows a limitation, because this potential does not define different sets of parameters for boron and nitrogen atoms. For the evaluation of mechanical/failure response of BC$_6$N nanosheet we thus only consider the Tersoff potential proposed by Kinarci *et al.* [110]. From the presented first-principles results (see Fig. 8b inset), we found that for the loading along the armchair direction the debonding initiates by the bond breakage between boron and carbon atoms. We therefore manually modified the original cutoff distance of 0.185 and 0.205 nm for boron-carbon bonds. We found that with a modified cutoff distance of 0.92 and 0.95 nm for boron-carbon bonds, best agreement can be reached. Even with the applied cutoff modifications, as it can be seen from the results shown in Fig. 9, the tensile strength of BC$_6$N monolayer along the armchair and zigzag directions are close, which is not consistent with first-principles findings. The deformation process of graphene and BC$_6$N nanosheets for the loading along the armchair directions with cutoff modified Tersoff potential at 300 K are depicted in Fig. 10. From the inset shown for the loading of BC$_6$N monolayer along the armchair direction, it is clear that the failure initiates by the breakage of boron-carbon bonds, which is in agreement with DFT results. Along the zigzag direction, the failure is however predicted to start between carbon-carbon bonds, while boron-nitrogen bonds are kept intact. It is clear that cutoff modified Tersoff potential exhibits remarkably enhanced accuracy for the modelling of the mechanical properties of graphene and BC$_6$N monolayers, but yet requires a further refinement to more accurately simulate mechanical/failure responses of these important 2D systems. The cutoff modified Tersoff potentials for the modeling of mechanical response of graphene and BC$_6$N nanosheets are included in the supplementary information document.

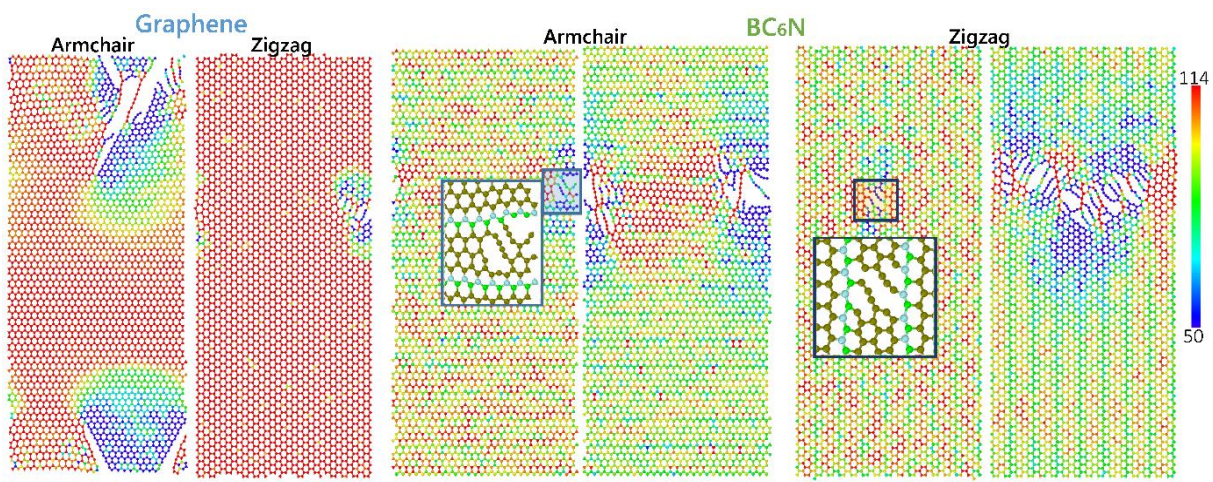

**Fig. 10**, Molecular dynamics predictions for the failure process of single-layer graphene and BC$_6$N along the armchair and zigzag loading directions at the room temperature. Contours depict uniaxial stresses with the unit of GPa, illustrated using the OVITO [117] package.

## 4. Concluding remarks

By conducting density functional theory calculations, we confirm that the most stable form of experimentally realized BC$_6$N layered material shows a rectangular unitcell. We moreover found that the BC$_6$N monolayer is a semiconductor with direct gaps of 0.71 and 1.19 eV according to the PBE and HSE06 methods, respectively. With anisotropic and excellent absorption of visible light, BC$_6$N monolayer can be a promising candidate for angle-dependent nanoelectronics and optoelectronic devices. The room temperature lattice thermal conductivity of BC$_6$N nanosheet along the zigzag and armchair directions are predicted to be 1890 and 1130 W/m.K, respectively, outperforming all other fabricated semiconducting layered materials. BC$_6$N monolayer shows ultrahigh mechanical properties, with the elastic modulus (tensile strength) of 945(111.1) and 920(87.5) GPa for the loading along the zigzag and armchair directions, respectively. We particularly developed classical molecular dynamic models for the evaluation of heat transport and mechanical properties of BC$_6$N nanomembranes, on the basis of first-principles findings. Acquired results not only shed light on the most stable atomic configuration of BC$_6$N nanosheet, but also confirm its outstanding electronic, optical, heat conduction and mechanical properties, extremely motivating for further theoretical and experimental endeavors.

## Acknowledgment

Author appreciates the funding by the Deutsche Forschungsgemeinschaft (DFG, German Research Foundation) under Germany's Excellence Strategy within the Cluster of Excellence PhoenixD (EXC 2122, Project ID 390833453). The VEGAS cluster at Bauhaus University of Weimar is acknowledged for providing the computational resources. Author does highly appreciate the insightful discussions with Prof. Shojaei from Persian Gulf University, Iran.

## Appendix A. Supplementary data

Supplementary data to this article are included after the References section.

Supplementary information

# Ultrahigh thermal conductivity and strength in direct-gap semiconducting graphene-like BC$_6$N: A first-principles and classical investigation

Bohayra Mortazavi

*Chair of Computational Science and Simulation Technology, Institute of Photonics, Department of Mathematics and Physics, Leibniz Universität Hannover, Appelstraße 11,30167 Hannover, Germany. Cluster of Excellence PhoenixD (Photonics, Optics, and Engineering–Innovation Across Disciplines), Gottfried Wilhelm Leibniz Universität Hannover, Hannover, Germany.*

Corresponding author: *bohayra.mortazavi@gmail.com;

1- Atomic structures in VASP-POSCAR
2- Uniaxial strain effect on the electronic band gap of BC$_6$N monolayer.
3- Tersoff potential to simulate thermal transport.
4- Tersoff potential to simulate mechanical properties.
5- Tersoff potential to simulate mechanical properties of graphene.

# 1-Atomic structures in VASP-POSCAR

**BC6N-Rec**
```
   1.00000000000000
     2.4744518981453592    0.0000000000000000    0.0000000000000000
     0.0000000000000000    8.6440881550933675    0.0000000000000000
     0.0000000000000000    0.0000000000000000   20.0000000000000000
   C    N    B
     6    1    1
Direct
  0.0017314987594688   0.3338202665501271   0.5000000000000000
  0.0017301921217978   0.8292037717567524   0.5000000000000000
  0.5017319466976033   0.4153341960121679   0.5000000000000000
  0.5017299562629134   0.9110638830093905   0.5000000000000000
  0.0017348553205920   0.6630182726002189   0.5000000000000000
  0.5017350064063208   0.5812418867076872   0.5000000000000000
  0.0017326355691765   0.1741098219222650   0.5000000000000000
  0.5017330094454309   0.0867039252103652   0.5000000000000000
```

**C3N**
```
   1.00000000000000
     4.8610886944767753    0.0000000000000000    0.0000000000000000
     2.4305443472383876    4.2098262994401487    0.0000000000000000
     0.0000000000000000    0.0000000000000000   20.0000000000000000
   C    N
     6    2
Direct
  0.1667028710785132   0.1667028710634995   0.5000000000000000
  0.6665942578379997   0.1667028710634995   0.5000000000000000
  0.1667028710785132   0.6665942578230002   0.5000000000000000
  0.8332971289164988   0.3334057421270060   0.5000000000000000
  0.3334057421570051   0.8332971288864996   0.5000000000000000
  0.8332971289164988   0.8332971288864996   0.5000000000000000
  0.6666666666650016   0.6666666666500021   0.5000000000000000
  0.3333333333300033   0.3333333333000041   0.5000000000000000
```

**BC3**
```
   1.00000000000000
     5.1731832464866967    0.0000000000000000    0.0000000000000000
     2.5865916227437409    4.4801081098096418    0.0000000000000000
     0.0000000000000000    0.0000000000000000   20.0000000000000000
   C    B
     6    2
Direct
  0.1587085081826842   0.1587085081689210   0.5000000000000000
  0.6825829836294020   0.1587085081679049   0.5000000000000000
  0.1587085081829116   0.6825829836131589   0.5000000000000000
  0.8412914918120933   0.3174170163368473   0.5000000000000000
  0.3174170163656029   0.8412914917820942   0.5000000000000000
  0.8412914918123207   0.8412914917810781   0.5000000000000000
  0.6666666666650016   0.6666666666500021   0.5000000000000000
  0.3333333333300033   0.3333333333000041   0.5000000000000000
```

**BC6N-1**
```
   1.00000000000000
     4.9779651560429246    0.0000000000000035    0.0000000000000000
    -2.4889825780231303    4.3110442842853924    0.0000000000000000
     0.0000000000000000    0.0000000000000000   20.0000000000000000
   C    N    B
   6    1    1
Direct
  0.3255055632853967  0.1627527816426877  0.5000000000000000
  0.1648550717673771  0.3297101435347685  0.5000000000000000
  0.8372472183573052  0.1627527816426877  0.5000000000000000
  0.1648550717673771  0.8351449282326229  0.5000000000000000
  0.8372472183573123  0.6744944367146175  0.5000000000000000
  0.6702898564652315  0.8351449282326229  0.5000000000000000
  0.3333333333333357  0.6666666666666643  0.5000000000000000
  0.6666666666666643  0.3333333333333357  0.5000000000000000
```

**BC6N-2**
```
   1.00000000000000
     4.9720634817238878   -0.0022396239068143    0.0000000000000000
    -2.4914361214161924    4.3107345704613484    0.0000000000000000
     0.0000000000000000    0.0000000000000000   20.0000000000000000
   C    N    B
   6    1    1
Direct
  0.3352903891286232  0.6676464513688245  0.5000000000000000
  0.1652659568547747  0.3295442957250927  0.5000000000000000
  0.8376425820591606  0.1615481110058781  0.5000000000000000
  0.1652660555316245  0.8357199436176259  0.5000000000000000
  0.8376435786259790  0.6761029015310669  0.5000000000000000
  0.6699686150179005  0.8349796994595309  0.5000000000000000
  0.3258129266297090  0.1629069189140750  0.5000000000000000
  0.6631098961522213  0.3315516783779060  0.5000000000000000
```

**BC6N-3**
```
   1.00000000000000
     4.9910202364654035    0.0000000000000000    0.0000000000000000
     2.4955101182327017    4.3038021741025236    0.0000000000000000
     0.0000000000000000    0.0000000000000000   20.0000000000000000
   C    N    B
   6    1    1
Direct
  0.1753930104879089  0.1640387525600531  0.5000000000000000
  0.6605682369320363  0.1640387525600531  0.5000000000000000
  0.1673125802920978  0.6653748393958168  0.5000000000000000
  0.3364467353542864  0.8337596548550792  0.5000000000000000
  0.8297936097506522  0.8337596548550792  0.5000000000000000
  0.6626531434508394  0.6746937130783266  0.5000000000000000
  0.3352010729312909  0.3295978540974218  0.5000000000000000
  0.8326316107809291  0.3347367783981667  0.5000000000000000
```

2- Uniaxial strain effect on the electronic band gap of BC$_6$N monolayer.

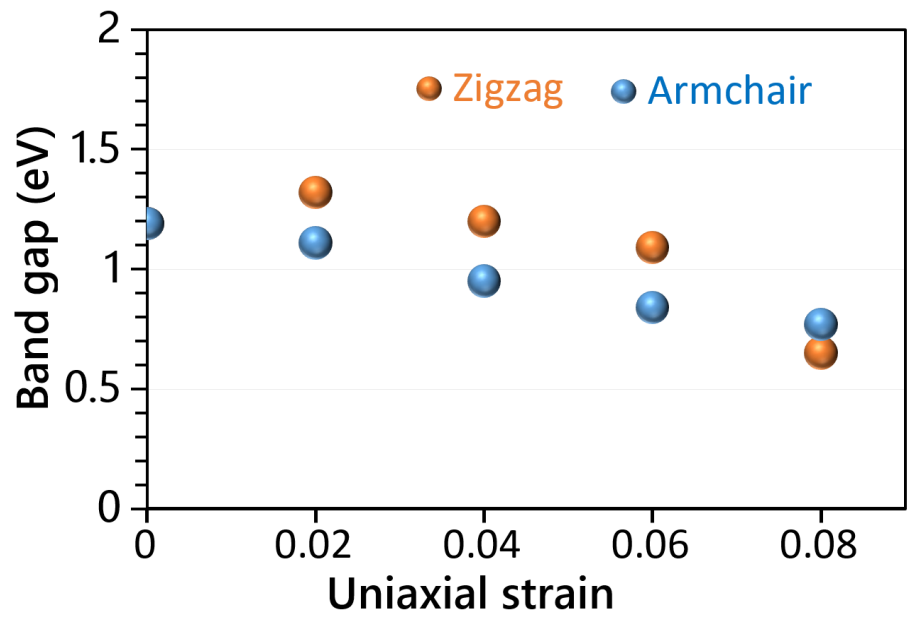

Fig. S1, Uniaxial strain effect on the electronic band gap of BC$_6$N monolayer.

### 3- Tersoff potential to simulate thermal transport.

```
# BC6N, Lindsay and Broido, C: PRB 81(2010),205441, B or N: PRB 84(2011),155421

# these entries are in LAMMPS "metal" units:
#   A,B = eV; lambda1,lambda2,lambda3 = 1/Angstroms; R,D = Angstroms
#   other quantities are unitless

# format of a single entry (one or more lines):
#   element 1, element 2, element 3,
#   m, gamma, lambda3, c, d, costheta0, n, beta, lambda2, B, R, D, lambda1, A

C   C   C   3.0 1.0 0.0 38049  4.3484   -0.930 .72751
            0.00000015724 2.2119  430.0   1.95   0.15   3.4879  1393.6

B   B   B   3.0 1.0 0.0 30692.4  4.7295 -0.98578 0.72674
             0.00000010239 2.2288  417.30 1.95  0.15   3.4664  1433.0

B   B   C   3.0 1.0 0.0 30692.4  4.7295 -0.98578 0.72674
             0.00000010239 2.2288  417.30 1.95  0.15   3.4664  1433.0

B   C   C   3.0 1.0 0.0 30692.4  4.7295 -0.98578 0.72674
             0.00000010239 2.22   425.30 1.95  0.15   3.477   1413.16

C   B   B   3.0 1.0 0.0 38049  4.3484   -0.930 .72751
             0.00000015724  2.22   425.30 1.95   0.15   3.477   1413.16

C   B   C   3.0 1.0 0.0 38049  4.3484   -0.930 .72751
             0.00000015724  2.22   425.30 1.95   0.15   3.477   1413.16

C   C   B   3.0 1.0 0.0 38049  4.3484   -0.930 .72751
             0.00000015724 2.2119  430.0   1.95   0.15   3.4879  1393.6

B   C   B   3.0 1.0 0.0 30692.4  4.7295 -0.98578 0.72674
             0.00000010239 2.22   425.30 1.95  0.15   3.477   1413
```

## 4- Tersoff potential to simulate mechanical properties.

```
# BC6N, C: PRB 81(2010),205441, B and N: PRB 86(2012), 115410
# Original by: Kinaci, Haskins, Sevik and Cagin, Phys Rev B, 86, 115410 (2012)
# Cutoff modified to reach tensile strength ~110 GPa for graphene at 300 K

# these entries are in LAMMPS "metal" units:
#   A,B = eV; lambda1,lambda2,lambda3 = 1/Angstroms; R,D = Angstroms
#   other quantities are unitless

# format of a single entry (one or more lines):
#   element 1, element 2, element 3,
#   m, gamma, lambda3, c, d, costheta0, n, beta, lambda2, B, R, D, lambda1, A

N       B       B       3.0 1.0 0.0 25000    4.3484 -0.89000 0.72751  1.25724e-7
2.199         340.00         1.95    0.05    3.568        1380.0
N       B       N       3.0 1.0 0.0 25000    4.3484 -0.89000 0.72751  1.25724e-7
2.199         340.00         1.95    0.05    3.568        1380.0
N       B       C       3.0 1.0 0.0 25000    4.3484 -0.89000 0.72751  1.25724e-7
2.199         340.00         1.95    0.05    3.568        1380.0

B       N       B       3.0 1.0 0.0 25000    4.3484 -0.89000 0.72751  1.25724e-7
2.199         340.00         1.95    0.05    3.568        1380.0
B       N       N       3.0 1.0 0.0 25000    4.3484 -0.89000 0.72751  1.25724e-7
2.199         340.00         1.95    0.05    3.568        1380.0
B       N       C       3.0 1.0 0.0 25000    4.3484 -0.89000 0.72751  1.25724e-7
2.199         340.00         1.95    0.05    3.568        1380.0

N       N       B       3.0 1.0 0.0 17.7959  5.9484  0.00000 0.6184432 0.019251
2.6272721     138.77866      2.0     0.1     2.8293093    128.86866
N       N       N       3.0 1.0 0.0 17.7959  5.9484  0.00000 0.6184432 0.019251
2.6272721     138.77866      2.0     0.1     2.8293093    128.86866
N       N       C       3.0 1.0 0.0 17.7959  5.9484  0.00000 0.6184432 0.019251
2.6272721     138.77866      2.0     0.1     2.8293093    128.86866

B       B       B       3.0 1.0 0.0 0.52629  0.001587 0.5    3.9929061  1.6e-6
2.0774982     43.132016      2.0     0.1     2.2372578    40.0520156
B       B       N       3.0 1.0 0.0 0.52629  0.001587 0.5    3.9929061  1.6e-6
2.0774982     43.132016      2.0     0.1     2.2372578    40.0520156
B       B       C       3.0 1.0 0.0 0.52629  0.001587 0.5    3.9929061  1.6e-6
2.0774982     43.132016      2.0     0.1     2.2372578    40.0520156

C       C       C       3.0 1.0 0.0 3.8049e4 4.3484 -0.93000 0.72751  1.5724e-7
2.2119  430.00    2.075    0.025    3.4879    1393.6
C       C       B       3.0 1.0 0.0 3.8049e4 4.3484 -0.93000 0.72751  1.5724e-7
2.2119  430.00    2.075    0.025    3.4879    1393.6
C       C       N       3.0 1.0 0.0 3.8049e4 4.3484 -0.93000 0.72751  1.5724e-7
2.2119  430.00    2.075    0.025    3.4879    1393.6

C       B       B       3.0 1.0 0.0 3.8049e4 4.3484 -0.93000 0.72751  1.5724e-7
2.2054  339.068910     1.935    0.015    3.5279    1386.78
C       B       N       3.0 1.0 0.0 3.8049e4 4.3484 -0.93000 0.72751  1.5724e-7
2.2054  339.068910     1.935    0.015    3.5279    1386.78
C       B       C       3.0 1.0 0.0 3.8049e4 4.3484 -0.93000 0.72751  1.5724e-7
2.2054  339.068910     1.935    0.015    3.5279    1386.78

C       N       B       3.0 1.0 0.0 3.8049e4 4.3484 -0.93000 0.72751  1.5724e-7
2.2054  387.575152     1.95    0.10    3.5279    1386.78
C       N       N       3.0 1.0 0.0 3.8049e4 4.3484 -0.93000 0.72751  1.5724e-7
2.2054  387.575152     1.95    0.10    3.5279    1386.78
C       N       C       3.0 1.0 0.0 3.8049e4 4.3484 -0.93000 0.72751  1.5724e-7
2.2054  387.575152     1.95    0.10    3.5279    1386.78

B       C       C       3.0 1.0 0.0 25000    4.3484 -0.89000 0.72751  1.25724e-7
2.2054  339.068910   1.935    0.015    3.5279    1386.78
B       C       B       3.0 1.0 0.0 25000    4.3484 -0.89000 0.72751  1.25724e-7
2.2054  339.068910   1.935    0.015    3.5279    1386.78
B       C       N       3.0 1.0 0.0 25000    4.3484 -0.89000 0.72751  1.25724e-7
2.2054  339.068910   1.935    0.015    3.5279    1386.78
```

```
N     C     C           3.0 1.0 0.0 25000    4.3484 -0.89000 0.72751 1.25724e-7
2.2054  387.575152      1.95    0.10    3.5279  1386.78
N     C     B           3.0 1.0 0.0 25000    4.3484 -0.89000 0.72751 1.25724e-7
2.2054  387.575152      1.95    0.10    3.5279  1386.78
N     C     N           3.0 1.0 0.0 25000    4.3484 -0.89000 0.72751 1.25724e-7
2.2054  387.575152      1.95    0.10    3.5279  1386.78
```

### 5- Tersoff potential to simulate mechanical properties of graphene.

```
# Lindsay and Broido, PRB 81(2010),205441
# Cutoff modified to reach tensile strength ~130 GPa for graphene at 300 K
# Reference: Carbon 103 (2016), 318-326, stable with time step = 0.25 fs

# these entries are in LAMMPS "metal" units:
#   A,B = eV; lambda1,lambda2,lambda3 = 1/Angstroms; R,D = Angstroms
#   other quantities are unitless

# format of a single entry (one or more lines):
#   element 1, element 2, element 3,
#   m, gamma, lambda3, c, d, costheta0, n, beta, lambda2, B, R, D, lambda1, A

C   C   C   3.0 1.0 0.0 38049  4.3484  -0.930 .72751
            0.00000015724 2.2119  430.0   2.05    0.05    3.4879  1393.6
```